\documentclass[prl,aps,showpacs,
preprint,
nofootinbib,
floatfix,superscriptaddress]{revtex4}
\usepackage{graphicx}
\usepackage{epsfig}
\usepackage{rotating}

\newcommand{\beq}{\begin{equation}}
\newcommand{\eeq}{\end{equation}}
\newcommand{\beqs}{\begin{eqnarray}}
\newcommand{\eeqs}{\end{eqnarray}}
\newcommand{\lsim}{\mathrel{\raisebox{-
.6ex}{$\stackrel{\textstyle<}{\sim}$}}}
\newcommand{\gsim}{\mathrel{\raisebox{-
.6ex}{$\stackrel{\textstyle>}{\sim}$}}}
\newcommand{\Tr}{{\rm Tr}}

\def\hbar{\hspace{0pt}\raisebox{1pt}{$-$} \hspace{-7pt} h}

\def\di{\mbox{d}}

\begin{document}
\title{Vector mesons from AdS/TC to the LHC}

\author{Maurizio Piai \thanks{email:piai@u.washington.edu }}
\affiliation{Department of Physics, University of Washington,
Seattle, WA 98195}

\date{April 17, 2007}

\begin{abstract}
With the use of the AdS-CFT dictionary, a five-dimensional 
effective description of dynamical electro-weak symmetry breaking with walking
behavior is constructed. The minimal  model contains only two new parameters, the 
confinement scale and the effective coupling of the new strong sector.
This parameter space is restricted by the precision electro-weak 
constraints and by the requirement that the five-dimensional coupling be 
perturbative (corresponding to the large-$N_T$ regime in four-dimensional language).
The lightest observable new states are a set of four nearly degenerate  
spin-1 states with the same quantum numbers as 
the standard-model electro-weak gauge bosons, and masses in the few TeV range.
Their decay rate is dominated by two-fermion final states.
The number of  $pp\rightarrow \mu^{+}\mu^{-}\,+\,X$ 
and $pp\rightarrow \mu^{+}\nu_{\mu}\,+\,X$  events
is studied as a function of the LHC integrated luminosity and of the two free parameters.
Discovery at the LHC is possible over a significant part of the allowed parameter space
up to masses of 4 TeV  already with 10 fm$^{-1}$ of integrated luminosity.
\end{abstract}


\maketitle

\section{Introduction}
The Large Hadron Collider (LHC) will  explore the physics responsible for
electro-weak symmetry breaking (EWSB) in the standard model (SM), and 
provide information about the specific mechanism responsible
for the generation of the mass of SM gauge bosons and fermions.
Dynamical EWSB, often referred to as technicolor (TC),
 assumes that a new strongly interacting
sector be responsible for the dynamical formation of a symmetry breaking 
condensate at the electro-weak scale~\cite{TC}, 
along the lines of  chiral symmetry breaking in QCD.
This idea provides a simple and elegant dynamical solution to the 
big hierarchy problem, i.~e. the huge separation between the 
electro-weak scale and the Planck scale.

Model-building in the TC context faces several problems,
which have over the years prevented from the formulation of a realistic and predictive model.
The intrinsic difficulty of treating analytically the strong dynamics
limits the extent to which a UV-complete model,
based on a non-abelian $SU(N_T)$ gauge theory, 
can be formulated and tested at low energies.
This fact enforces the use of an effective field theory (EFT) description 
when comparing to the experiments: the electro-weak chiral Lagrangian (EW-$\chi$L)~\cite{EWCL}.
This has to be completed with a mechanism for the generation of mass for the SM fermions,
generically referred to as extended technicolor (ETC), which connects the SM fermions
and the symmetry-breaking condensates~\cite{ETC} via higher-order operators
arising from new interactions.

Many non-trivial properties of this EFT are known experimentally.
The contribution of dimension-6 operators in the EW-$\chi$L  to the $\hat{S}$ and $\hat{T}$ 
parameters~\cite{PT} have to be at the most in the few $\times \,10^{-3}$ range~\cite{Barbieri}, 
in contrast with much larger expectations from naif dimensional analysis (NDA).
The typical scale of ETC must be in the $4-5$ TeV range or above,
which produces some tension with the requirement that ETC be efficient enough
as to produce the observed large top mass.
From direct and indirect searches, the low-energy spectrum
should not contain any light pseudo-Nambu-Goldstone boson (pNGB), with masses
in the few GeV range or below, which are a generic prediction of many TC models.
Finally, the ETC sector must produce the observed pattern of masses
and mixing angles for the SM fermions, while adequately suppressing 
new physics contributions to flavor changing neutral current
(FCNC) processes.

In the context of  EW-$\chi$L with generic ETC couplings, 
all of the above requirements can be 
satisfied simply by tuning to unnaturally small (or unnaturally large) 
values some of the coefficients
in the EFT, and by assuming that the unknown UV physics 
(i.~e. the underlying strong dynamics) be responsible for these
large deviations from the natural expectations. 
In particular, it has long ago been suggested that a large 
cut-off scale might be dynamically separated from the electro-weak scale,
if the underlying strong dynamics is approximately conformal over a 
finite energy window ({\it walking} behavior~\cite{walking}). 
This can suppress the new physics contributions to 
electro-weak precision parameters~\cite{AS} and FCNC currents, but not 
necessarily the top mass, provided the chiral condensate has a large anomalous dimension 
(for instance if the condensate has scaling dimension $d=2$).
Also, semi-realistic UV complete models of ETC have been proposed,
suggesting that identifying the lightest ETC scale with  this $4-5$ TeV cut-off
is compatible with the experimental properties of the SM mass matrices
and with FCNC constraints~\cite{APS}.

However, in the absence of a reliable, first principle computation of the
effective Lagrangian coefficients,
the choice of renouncing the criterion 
of naturalness, and hence of allowing for large departures from their NDA 
estimates of the very big (potentially infinite) number of  free parameters in the
effective Lagrangian, leads to  loss of predictive power.
It is hence necessary to extend the EFT validity up to higher scales.
The natural direction in which to carry this program is to extend the field content
to include not only the SM gauge bosons, but also some of the lightest composite states
emerging at the confinement scale from the strong dynamics. 
Such an approach has
been followed over the years in the context of QCD, 
starting with the concept of vector dominance,
developed with the techniques of hidden local symmetry~\cite{hidden} and,
more recently, of deconstruction~\cite{deconstructQCD},
and applied to the description of dynamical EWSB~\cite{deconstructTC}. 
Though elegant,  simple and very useful, 
this approach does not solve the problem of the proliferation of 
free parameters because the number of independent operators in the 
effective Lagrangian grows very fast with the number of new fields.
A more fundamental guiding principle is necessary. 

More recently, in the context of string theory, compelling evidence has been collected
confirming that some special super-Yang-Mills (SYM) strongly coupled, 
conformal field theories (CFT), in the regime of large number $N_c$ of fundamental colors,
admit a dual description in terms of a weakly coupled, higher-dimensional gravitational theory
with Anti-de Sitter (AdS) background.  This is the AdS/CFT correspondence~\cite{AdS/CFT}.
The relative simplicity and enormous power of the computational techniques provided
by this correspondence suggest the idea that by perturbing and deforming 
in a controlled way
the pure AdS background it is possible to explore a large class of  
weakly-coupled extra-dimension theories,
that is thought to provide a dual description for a much larger class of strongly-coupled four-dimensional systems, in which both supersymmetry and conformal symmetry are (softly) broken.
Simple phenomenological constructions are possible~\cite{pheno} along these lines.
This idea has been applied with interesting results to the construction of 
an EFT description of QCD valid well beyond the
reach of the traditional chiral Lagrangian~\cite{AdS/QCD}.

The AdS/CFT correspondence is the guiding principle 
that might prove crucial in the construction 
of a viable and predictive EFT description of dynamical electro-weak 
symmetry breaking~\cite{AdS/TC}~\cite{MP}.
 I first  briefly review the construction of the most minimal such 
EFT description of dynamical EWSB, based on the indications
collected by experimental data.
The model contains only two new free parameters, and the parameter space is  
restricted by electro-weak precision data. 
In the main body of the paper I discuss the  phenomenology
of the spin-1 composite states (techni-$\rho$). Their decay rates
are dominated, in the large-$N_T$ regime, 
by two-body decays into SM fermions. An upper bound on the strength of the
relevant coupling is found,
hence allowing for a  controllable, perturbative study
of the LHC discovery reach.
A simple analysis of the production and detection at the LHC of the spin-1 states
is then performed, in order to quantify the reach of the experimental project.
In the  $pp\rightarrow \ell\ell^{\prime}\,+\,X$ processes, for which the SM background
is  very low, the uncertainties related to jet and hadronic physics 
are not crucial  and the detector efficiency is good, a very large portion
of the parameter space will be explored within the first few years.

Before beginning to construct and study the model,
a comment is due, about the nature of the two aforementioned free parameters.
If the underlying dynamics is that of some $SU(N_T)$ gauge theory
with some given field content, it should be possible to compute from
first principles the scale of confinement, the scale of electro-weak symmetry
breaking, the scaling dimension of the chiral condensate and 
the strength of the effective couplings of the composite states,
besides proving that the theory admits an approximate IR fixed point,
leading to quasi-conformal behavior near the confinement scale.
Unfortunately, this cannot be done  systematically and reliably 
 with existing methods~\cite{Shrock}, and  hence
all of these quantities are going to be implemented in the model as free parameters.

The electro-weak symmetry breaking scale is fixed experimentally through the
measured value of the Fermi constant $G_F$. The scaling dimension $d=2$ is chosen,
somewhat arbitrarily, in order to reproduce the scaling arguments
of walking technicolor, leading to a large enough  mass for the top quark,
 while preserving universality of the weak couplings.
The confinement scale (related to $L_1$ defined in the body of the paper),
and the strength of the effective couplings (see $\varepsilon^2$ later on),
cannot be decided a priori, and are treated as completely free parameters.
As will be clear, a limit on the former is given by  precision electro-weak physics,
on the latter by the requirement that the perturbative expansion
used here (and related to the large-$N_T$ expansion), be valid.
Experimentally, these two parameters are related to the
mass $M$ of the excited spin-1 states, and to their relative coupling $R$
 to the SM currents, measured in units of the SM gauge coupling,
 both which can be extracted from $pp\rightarrow \ell\ell^{\prime}\,+\,X$ processes.
Remarkably, a very large part of the parameter space identified in this way 
is going to be testable at the LHC even in its early stages.

\section{The Model.}

A summary of  the minimal requirements for a viable model of dynamical EWSB
in four dimensions allows to identify the properties of the five dimensional dual description. 
There must be a new strong sector possessing the global symmetry
$SU(2)_L\times U(1)_Y$ of the standard model.
The new interaction must confine, and a symmetry breaking condensate
must form. The (weak) gauging of the global symmetry of the strong sector
gives the massive SM gauge bosons and the photon.
In order for higher order operators contributing to precision 
electro-weak parameters and FCNC  currents to be suppressed by a large enough scale,
the strong sector has to be close to conformal over the energy range 
between the confinement scale and a  larger ETC scale of $4-5$ TeV.
The top mass can be accommodated by assuming that the chiral condensate
has scaling dimension $d=2$. All the SM fermions are elementary, and do not carry
quantum numbers of the new strong interactions, hence ensuring 
universality of the electro-weak gauge coupling.

The (quasi-conformal) energy window just above the electro-weak scale
is described by a slice of $AdS_5$, 
i.e. by a five-dimensional space-time containing a warped gravity 
background given by the metric:
\beqs
\di s^2 &=& \left(\frac{L}{z}\right)^{2}\left( \eta_{\mu\nu}\di x^{\mu}\di x^{\nu}\,-\,\di z^2\right)\,,
\eeqs
where $x^{\mu}$ are four-dimensional coordinates, $\eta_{\mu\nu}$ the Minkoski 
metric with signature $(+,-,-,-)$, and $z$ is the extra (warped) dimension.
The dimensionful parameter $L$ is the $AdS_5$ curvature, and sets the 
overall scale of the model.
Conformal symmetry is broken by  the boundaries 
\beqs
L_0\,<\,z\,<\,L_1\,,
\eeqs
with $L_0>L$, where $L_0$ and $L_1$ correspond to the UV and IR cut-offs of the
conformal theory, i.~e. to the ETC scale and to the confinement scale respectively.

The field content in the bulk of the five-dimensional  model 
consists of a  complex scalar $\Phi$ transforming 
as a $(2,1/2)$ of the gauged  $SU(2)_L\times U(1)_Y$.
The generator of $SU(2)_L$ are $T_i=\tau_i/2$ with $\tau_i$ the 
Pauli matrices.

The bulk action for $\Phi$ and the gauge bosons $W = W_i T_i$ of $SU(2)_L$  
and $B$ of $U(1)_Y$ is 
\beqs
{\cal S}_{5} &=& \int\di^4 x \int_{L_0}^{L_1}\di z\,\sqrt{G}\left[\frac{}{}
\left(G^{MN}(D_M\Phi)^{\dagger} D_N\Phi -M^2|\Phi|^2\right)\,\nonumber\right.\\
&&\left. \left(-\frac{1}{2}\Tr\left(W_{MN}W_{RS}\right)-\frac{1}{4}B_{MN}B_{RS}\right)G^{MR}G^{NS}\right]\,,
\eeqs
and the boundary terms are 
\beqs
{\cal S}_{4} &=& \int\di^4 x \int_{L_0}^{L_1}\di z \,\sqrt{G}\left[ \frac{}{}\delta(z-L_0)\,D\right.\\
&&\left[-\frac{1}{2}\Tr\left[W_{\mu\nu}W_{\rho\sigma}\right]-\frac{1}{4}B_{\mu\nu}B_{\rho\sigma}\right]
G^{\mu\rho}G^{\nu\sigma}\nonumber\\
&& -\delta(z-L_0)\,2\lambda_0 \left(|\Phi|^2-\frac{\mbox{v}_0^2}{2}\right)^2\nonumber\\
&&\left. -\delta(z-L_1)\,2\lambda_1 \left(|\Phi|^2-\frac{\mbox{v}_1^2}{2}\right)^2\right]\,,
\eeqs
where the covariant derivative is given by
\beqs
D_M\Phi &=& \partial_M \Phi + i (g W_M \Phi +\frac{1}{2} g^{\prime}B_M  \Phi )\,,\nonumber
\eeqs
and where the Yang-Mills action is written in terms of the antisymmetric 
field-strength tensors $W_{\mu\nu}$ and $B_{\mu\nu}$.
In the action, $M^2$ is a bulk mass term for the scalar, 
and $g$ and $g^{\prime}$ are
the (dimensionful) gauge couplings in five-dimensions.

Without loss of generality:
\beqs
\langle \Phi \rangle &=& \frac{\mbox{v}(z)}{\sqrt{2}}\left(\begin{array}{c}
0\cr 1\end{array}\right)\,.
\eeqs
The localized potentials produce the $z$-dependent vacuum expectation value (VEV)
 of the scalar field that induces 
electro-weak symmetry breaking.  For $M^2=-4/L^2$, in the $\lambda_i\rightarrow +\infty$ limit,
 (in which the transverse degrees of freedom become infinitely massive and decouple from the spectrum) the bulk equation of the motion 
\beqs
\partial_z\left(\frac{L^3}{z^3}\partial_z{\rm v}\right)-\frac{L^5}{z^5}M^2{\rm v}=0\,,
\eeqs
admits the solution
\beqs
\mbox{v}(z)&=&\frac{\mbox{v}_1}{L_1^2}z^2\,=\,\frac{\mbox{v}_0}{L_0^2}z^2\,,
\eeqs
by appropriately choosing  $\mbox{v}_{0}/\mbox{v}_1$, 
so as to describe a chiral condensate of dimension $d=2$.

The localized kinetic terms for the gauge bosons
are required by holographic renormalization~\cite{HR} 
in order to retain a finite SM gauge coupling
in the $L_0\rightarrow 0$ limit, and renormalizing the otherwise divergent
 kinetic terms of the SM gauge bosons.
This procedure ensures that  the SM gauge couplings
be independent of the strength of the bulk coupling, 
and that all physical quantities be independent of any 
UV-sensitive details, such as the precise value of $L_0$ 
and the UV-brane dynamics.

The SM fermions, and hence the SM currents, are localized at the UV boundary.
They do not feel the strong interactions, and their mass
is induced by localized, Yukawa-like dimension-5 couplings 
to the symmetry-breaking background.
An alternative way, not explored here, of enhancing the top mass consists of 
assuming that the top (or possibly the whole third family)
takes part directly in the strong dynamics,  as in topcolor models~\cite{topcolor}.
In this case the top would be composite,  
and hence would propagate in the fifth dimension, as happens 
in the higgsless context~\cite{higgsless}
and in several recent composite Higgs models~\cite{composite}. While
admissible, these models violate tree-level universality of the gauge interactions,
hence adding to the difficulties connected with precision electro-weak 
and FCNC constraints,
and lead to the introduction of a much larger set of free parameters in the 
SM fermion sector.
Several other examples of viable, but  non-minimal, choices, 
not discussed in the following,
might be the addition of localized kinetic terms for 
the scalar, or the choice of inducing non trivial mixing in the localized kinetic terms
for the gauge bosons, or delocalizing the first two family fermions,
all of which would affect the parameters $\hat{S}$,
or the choice of gauging the whole $SU(2)_R$ in the bulk, so as to 
have a custodial symmetry for $\hat{T}$. 

\section{Electro-weak Phenomenology.}

I discuss only the spin-1 sector of the model, in unitary gauge.
The spin-0 sector consists of the transverse degrees of freedom of 
$\Phi$, integrated out  in the $\lambda_i\rightarrow \infty$
limit, and the pseudo-scalar sector, which does not contain a zero mode,
and is neglected in this paper for the sake of simplicity.

I define, as usual:
\beqs
V^{M}&\equiv & \frac{g^{\prime}W_3^{M}+gB^{M}}{\sqrt{g^2+g^{\prime\,2}}}\,,\\
A^{M}&\equiv & \frac{g W_3^{M}-g^{\prime}B^{M}}{\sqrt{g^2+g^{\prime\,2}}}\,,
\eeqs
so that  the massless 
mode of $V^{\mu}$ is the photon, and the lightest
mode of $A^{\mu}$ is the $Z$ boson.

After Fourier transformation in the four-dimensional Minkoski coordinates:
\beqs
A^{\mu}(q,z)&\equiv&A^{\mu}(q)v_Z(z,q)\,,
\eeqs
and analogous for $W_{1,2}$ and $V$, 
where $q=\sqrt{q^2}$ is the four-dimensional momentum.
In the limit in which one neglects 
the gauge coupling, and hence cubic and quartic self-interactions, the
bulk equations are:
\beqs
\partial_z\frac{L}{z}\partial_z v_i-\mu^4_{i} L z  v_i&=&-q^{2}\frac{L}{z}v_i\,,
\eeqs
where $i=v,Z,W$, with $\mu_v=0$,  $\mu^4_W=1/4g^2\mbox{v}_0^2/L^2$ and 
$\mu^4_Z=1/4(g^2+g^{\prime 2})\mbox{v}_0^2/L^2$.
This approximation, and the approximation of restricting all the analysis
at the tree-level, are valid provided the effective coupling $g/\sqrt{L}\lsim O(1)$,
which corresponds to  the large-$N_T$ limit.

The bulk equations can be solved exactly.  
Substituting the solutions in the action, the bulk part of the action vanishes
identically. Choosing Neumann boundary conditions in the IR
leaves a non-vanishing UV-localized action, that can be written
as
\beqs
{\cal L}&=&-\frac{1}{2}A_{\mu}^i\,\pi_{i,j} P^{\mu\nu}A^{j}_{\nu}\,,
\eeqs
where $P^{\mu\nu}\equiv\eta^{\mu\nu}-q^{\mu}q^{\nu}/q^2$,
and where $i=B,W_a$.
The matrix of the polarizations $\pi_{i,j}(q^2)$
of the SM gauge bosons can be written in terms of the 
action evaluated at the UV boundary:
\begin{widetext}
\beqs
\frac{\pi_{+}}{{\cal N}^2}&=&Dq^2+\frac{\partial_zv_{W}}{v_{W}}(q^2,L_0)\,,
\\
\frac{\pi_{BB}}{{\cal N}^2}&=&
Dq^2 +\frac{g^2}{g^2+g^{\prime\,2}}\frac{\partial_zv_{v}}{v_v}(q^2,L_0)
+\frac{g^{\prime\,2}}{g^2+g^{\prime\,2}}\frac{\partial_zv_{Z}}{v_Z}(q^2,L_0)\,,
\\
\frac{\pi_{WB}}{{\cal N}^2}&=&
\frac{g g^{\prime}}{g^2+g^{\prime\,2}}\left(\frac{\partial_zv_{v}}{v_v}(q^2,L_0)
-\frac{\partial_zv_{Z}}{v_Z}(q^2,L_0)\right)\,,\\
\frac{\pi_{WW}}{{\cal N}^2}&=&
Dq^2 +
\frac{g^{\prime\,2}}{g^2+g^{\prime\,2}}\frac{\partial_zv_{v}}{v_v}(q^2,L_0)
+\frac{g^{2}}{g^2+g^{\prime\,2}}\frac{\partial_zv_{Z}}{v_Z}(q^2,L_0)\,,
\eeqs
\end{widetext}
where ${\cal N}$ is 
chosen to produce canonical kinetic terms in the
limit in which the heavy resonances  decouple.
The precision electro-weak parameters are defined as
\beqs
\hat{S}&\equiv&\frac{g_4}{g_4^{\prime}}\,\pi_{WB}^{\prime}(0)\,,\\
\hat{T}&\equiv&\frac{1}{M_W^2}\left(\pi_{WW}(0)-\pi_{+}(0)\right)\,,
\eeqs
where  $g_4^{(\prime)}$
 are the (dimensionless) 
gauge couplings of the SM in four-dimensions,
and where $\pi^{\prime}\equiv \di \pi/\di q^2$.

Taking for simplicity $L_0\rightarrow L$, and expanding  for small  $L_0\rightarrow 0$,
from~\cite{MP}:
\begin{widetext}
\beqs
\frac{\partial_zv_{v}}{v_v}(q^2,L_0)&=&q^2L_0\left(\frac{\pi}{2}
\frac{Y_0(qL_1)}{J_0(qL_1)}-\left(\gamma_E+\ln \frac{qL_0}{2} \right)\right)\,,\\
\frac{\partial_zv_{Z}}{v_Z}(q^2,L_0)&=&
L_0\left\{
\mu_Z^2\,-\,q^2\left[
\gamma_E+\ln(\mu_Z L_0)+\frac{1}{2}\psi\left(-\frac{q^2}{4\mu_Z^2}\right)-\frac{c_2}{2c_1}\Gamma\left(-\frac{q^2}{4\mu_Z^2}\right)
\right]\right\}\,,
\eeqs
where, having  imposed Neumann boundary conditions in the IR, 
\beqs
c_1&=&2L\left(-1+\frac{q^2}{4\mu_Z^2},\mu_Z^2L_1^2\right)+L\left(\frac{q^2}{4\mu_Z^2},-1,\mu_Z^2L_1^2\right)\,,\\
c_2&=&-U\left(-\frac{q^2}{4\mu_Z^2},0,\mu_Z^2L_1^2\right)+\frac{q^2}{2\mu_Z^2}
U\left(1-\frac{q^2}{4\mu_Z^2},1,\mu_Z^2L_1^2\right)\,,
\eeqs
\end{widetext}
and where $\partial_zv_{W}/v_{W}=
\partial_zv_{Z}/v_{Z}(\mu_Z\rightarrow \mu_W)$. 

The  localized counterterm 
\beqs
D&=&L_0\left(\ln\frac{L_0}{L_1}+\frac{1}{\varepsilon^2}\right)\,
\eeqs
cancels the logarithmic divergences, and for ${\cal N}^2=\varepsilon^2/L_0$ 
all the dependence on $L_0$  disappears (at leading order in $L_0$), 
the limit $L_0\rightarrow 0$ can be taken,
and the model is renormalized, with finite (dimensionless) SM couplings
$g_4^{(\prime)\,2}= \varepsilon^{2}g^{(\prime)\,2}/L$.

Expanding for $\mu^2_ZL_1^2\ll1$, 
\beqs
\hat{S}&=&\varepsilon^2
\frac{1}{2e}\mu_W^4L_1^4\,,\\
\hat{T}&=&\frac{\varepsilon^2}{M_W^2}\left(\mu_W^2\tanh\frac{\mu_W^2L_1^2}{2}-
\frac{\mu_W^4}{\mu_Z^2}\tanh\frac{\mu_Z^2L_1^2}{2}\right)\,\\
&\simeq&\frac{\varepsilon^2}{M_W^2}\frac{\mu_W^4L_1^6}{24}(\mu_Z^4-\mu_W^4)\,,
\eeqs
with $e\simeq 2.7$.

The mass of the  $W$ gauge boson is approximately given by
\beqs
M_W^2&\simeq&\varepsilon^2\left(\mu_W^2\tanh\frac{\mu_W^2L_1^2}{2}\right)\,\simeq\,\frac{1}{2}\varepsilon^2\mu_W^4L_1^2\,,
\eeqs
while $M_Z^2\simeq (g^2+g^{\prime\,2})/g^2M_W^2$.

Substituting  in
the precision parameters yields:
\beqs
\hat{S}&\simeq&\frac{1}{e}M_W^2L_1^2
,\\
\hat{T}&\simeq&\frac{M_Z^2-M_W^2}{6\varepsilon^{2}}L_1^2
\,=\, \frac{e}{6\varepsilon^2}\frac{M_Z^2-M_W^2}{M_W^2}\,\hat{S}\,.
\eeqs

I take as indicative of the experimentally allowed ranges
(at the $3\sigma$ level):
\beqs
\hat{S}_{exp}&=& (-0.9\pm 3.9) \times 10^{-3}\,,\\
\hat{T}_{exp}&=& (2.0\pm3.0) \times 10^{-3}\,,
\eeqs
from~\cite{Barbieri}. These bounds  are extrapolated to
the case of a Higgs boson with mass of $800$ GeV. The  comparison has to be 
done with some caution, since  the one-loop level SM analysis used
in the extraction of the bounds is not reliable for a heavy, strongly coupled Higgs,
the mass of which is not controllable in this model.

For reasonable values of $\varepsilon > 1/2$, such that the five-dimensional  tree-level 
analysis performed here holds, the bounds 
on $\hat{T}$ have no significant effect in restricting the parameter space of the model.
The implementation of a custodial symmetry is unnecessary.
Form the approximate expression for $\hat{S}$ comes the limit on the
confinement scale $L_1$ of the model:
\beqs
L_1^2&\lsim&e\,\frac{ \sup\left(\hat{S}_{exp}\right)}{M_W^2}\,\simeq\,\frac{1}{(890\, {\rm GeV})^2}\,.
\eeqs

\section{Vectorial modes.}

I focus on the lightest and next to lightest modes of the gauge bosons, neglecting
completely all the other modes.
The lightest modes are the SM gauge bosons. 
The next excited states are the techni-$\rho$ resonances
 which I indicate from now on as $\gamma^{\prime}$, $Z^{\prime}$ and $W^{\prime}$,
because they couple to the same
 SM currents as the photon, the $W$ and the $Z$ bosons.
Their production (and decay) rates depend only on the
masses and on the coupling to the SM currents,
which are computed explicitly in this section.

The solutions of the bulk equations, after imposing the Neumann 
boundary conditions in the IR, read:
\beqs
v_v(z,q)&=&c_v(q) \,z \left(J_0(qL_1)Y_1(qz)-Y_0(qL_1)J_1(qz)\right)\,,\\
v_a(z,q)&=&c_a(q)\,
e^{\frac{-\mu_Z^2z^2}{2}}\left[c_1(q)U\left(-\frac{q^2}{4\mu_Z^2},0,\mu_Z^2 z^2\right)\,+\,
c_2(q)L\left(\frac{q^2}{4\mu_Z^2},-1,\mu_Z^2z^2\right)\right]\,,\\
v_{+}(z,q)&=&c_{+}(q)\,
e^{\frac{-\mu_W^2z^2}{2}}\left[c_1^{\prime}(q)U\left(-\frac{q^2}{4\mu_W^2},0,\mu_W^2 z^2\right)\,+\,
c_2^{\prime}(q)L\left(\frac{q^2}{4\mu_W^2},-1,\mu_W^2z^2\right)\right]\,,
\eeqs
where $i=v,a,+$, $c_i$ are normalization constants,  determined,  at finite $L_0$,  by
\beqs
1&=&\int_{L_0}^{L_1}\di z \frac{L}{z} |v_i(z,q)|^2\left(\frac{}{}1+D\delta(z-L_0)\right)\,,
\eeqs 
with
\beqs
D&=&L_0\left(\ln\frac{L_0}{L_1}+\frac{1}{\varepsilon^2}\right)\,,
\eeqs   
while
$c_{1,2}^{\prime}\equiv c_{1,2}(\mu_Z\rightarrow \mu_W)$.

The masses of the techni-$\rho$  are obtained by solving:
 \beqs
 D M_{\gamma^{\prime}}^2\,+\, \frac{\partial_z v_v(L_0,M_{\gamma^{\prime}})}{v_v(L_0,M_{\gamma^{\prime}}) }&=&0\,,\\
 D M_{W^{\prime}}^2 \,+\, \frac{\partial_z v_{+}(L_0,M_{W^{\prime}}) }{v_{+}(L_0,M_{W^{\prime}})}&
   =&0\,,\\
 D M_{Z^{\prime}}^2 \,+\, \frac{\partial_z v_{a}(L_0,M_{Z^{\prime}}) }{v_{a}(L_0,M_{Z^{\prime}})}&
   =&0\,.
 \eeqs
 
\begin{figure*}[ht!]
\begin{center}
\includegraphics[width=0.8\linewidth]{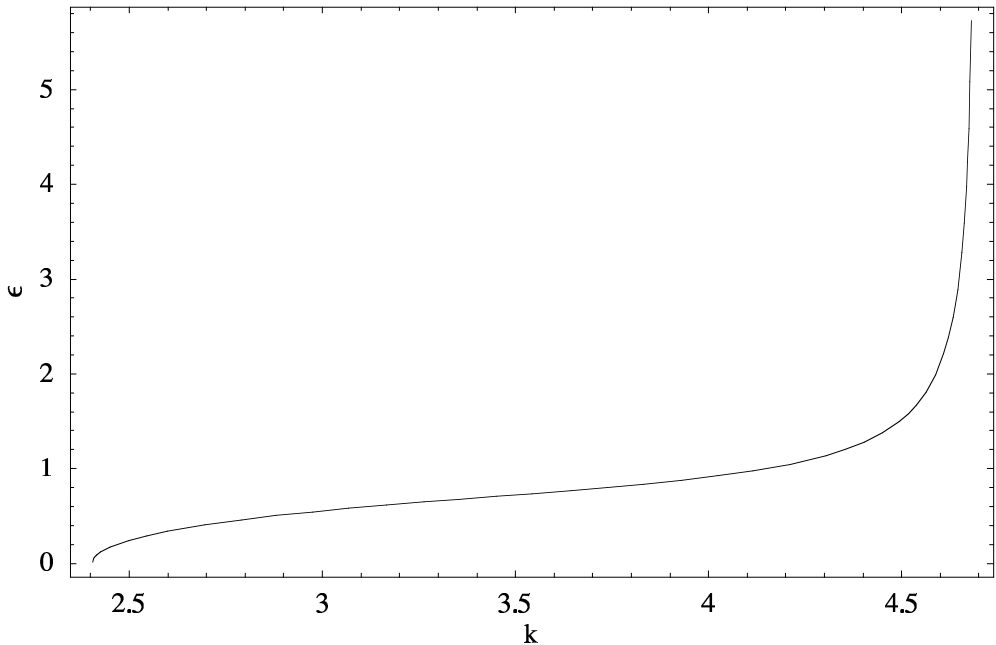}
\caption{The dependence of $\varepsilon$ on $k$, the parameter determining the mass of $\gamma^{\prime}$.\label{Fig:mass}}
\end{center}
\end{figure*}

The mode corresponding to the excited  photon has a mass
\beqs
M_{\gamma^{\prime}}&=&\frac{k}{L_1}\,,
\eeqs
where $k_0<k<4.7$ is a monotonically growing  function of $\varepsilon$,
with $k_0\simeq 2.4$ being the first zero of $J_0(z)$.
The (exact) relation between $\varepsilon$ and  $k=M_{\gamma^{\prime}}L_1$ is given by
\beqs
\frac{1}{\varepsilon^2}&=&\gamma_E+\ln\frac{k}{2}
-\frac{\pi}{2}\frac{Y_0(k)}{J_0(k)}\,,
\eeqs
and is illustrated in Fig.~\ref{Fig:mass}.
The bounds on $\hat{S}$ imply that  the mass of the techni-$\rho$ is much larger than 
the mass of the SM gauge bosons. In turns, this implies that, for all practical purposes,
the masses of the $Z^{\prime}$ and $W^{\prime}$ fields are degenerate
with $M_{\gamma^{\prime}}$. 

Notice that the  smallest values of $k\simeq k_0$ correspond to small values of $\varepsilon$,
and the largest value of $k$ is reached for asymptotically large values of $\varepsilon$.
Very small values of $\varepsilon$ correspond to the gauging of the global symmetry
of the system being very small, in comparison with the effective coupling of the composite states.
In this case, the wave functions of the techni-$\rho$ states are approximately given by the normalizable solution to the bulk equations, that vanishes for $z\rightarrow 0$. 
Going to larger values of $\varepsilon$ requires a finite, increasing value of 
the wave function at the UV-boundary, and hence the localized term selects a linear combination
of normalizable and non-normalizable solutions, with increasing contribution
from the latter. As a result, the mass is a growing function of $\varepsilon$,
but the growth
saturates at large $\varepsilon$ toward a constant.
This somewhat peculiar behavior has to do with the way the limits are taken 
in the present context: the SM gauge couplings (known experimentally) are
kept fixed while going to large-$N_T$, so that the gauging of the 
global symmetry, although always weak and controllable, 
becomes stronger than the effective coupling of the strong sector itself, and hence 
its effect is not just a perturbation of the expectations deduced by analyzing 
the large-$N_T$ limit of a generic strongly coupled model 
with a global symmetry.

The couplings to the electro-magnetic, neutral and charged weak currents
of the techni-$\rho$'s are related to the SM gauge boson couplings by
the $L_0\rightarrow 0$ limit of:
\beqs
e_4^{\prime}&=&e_4\,\frac{v_v(L_0,M_{\gamma^{\prime}})}{v_v(L_0,0)}\,,\\
g_{c.c.}^{\prime}&=&g_{c.c.}\,\frac{v_{+}(L_0,M_{W^{\prime}})}{v_{+}(L_0,M_W)}\,,\\
g_{n.c.}^{\prime}&=&g_{n.c.}\,\frac{v_{a}(L_0,M_{Z^{\prime}})}{v_{a}(L_0,M_Z)}\,.
\eeqs

\begin{figure*}[ht!]
\begin{center}
\includegraphics[width=0.8\linewidth]{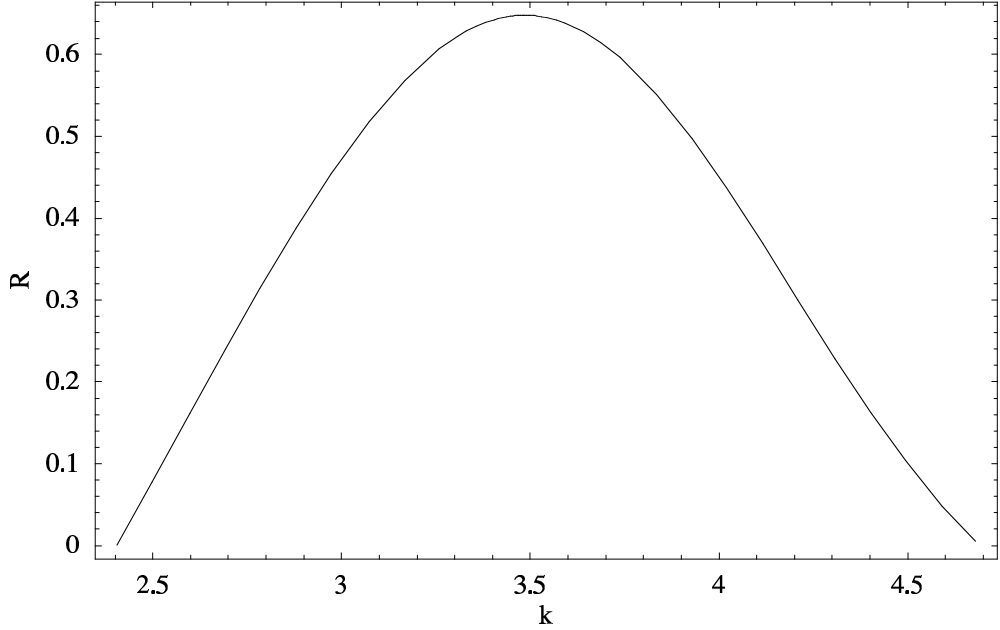}
\caption{The ratio $R=|e_4^{\prime}/e_4|^2$ between the couplings to the 
electro-magnetic current of the $\gamma^{\prime}$ boson and the photon,
as a function of $k$.\label{Fig:coupling}}
\end{center}
\end{figure*}

For the electro-magnetic case, it is  possible to explicitly write
(making use of the relation between
 $k$ and $\varepsilon$)
the dependence on $k$ of 
the ratio of couplings to the SM fermions as 
illustrated by Figure~\ref{Fig:coupling}:
\beqs
\frac{1}{R}\,\equiv\,\left(\frac{e_4}{e^{\prime}_4}\right)^2&=&
\frac{\left(\pi ^2 \left(Y_0(k)
   Y_2(k)-Y_1(k)^2\right) k^2+4\right)
   J_0(k)^2+\pi  Y_0(k) \left(\pi  J_2(k)
   Y_0(k) k^2+4\right) J_0(k)}{4 J_0(k) \left(\pi  Y_0(k)-2
   J_0(k) \left(\log \left(\frac{k}{2}\right)+\gamma
   \right)\right)}\nonumber\\
   &&-\frac{2 \pi ^{3/2}
   Y_0(k) G_{2,4}^{2,1}\left(k^2\left|
\begin{array}{c}
 1,\frac{3}{2} \\
 1,2,0,0
\end{array}\right.
\right) J_0(k)-k^2 \pi ^2 J_1(k)^2
   Y_0(k)^2}{4 J_0(k) \left(\pi  Y_0(k)-2
   J_0(k) \left(\log \left(\frac{k}{2}\right)+\gamma
   \right)\right)}\,.
\eeqs
Again, this quantity is only marginally sensitive to electro-weak symmetry breaking, and 
hence $g_{c.c.}^{\prime}/g_{c.c.}\simeq g_{n.c.}^{\prime}/g_{n.c.}\simeq e_4^{\prime}/e_4$.

Some comments are in order.
First of all, notice the absolute upper bound on $R\lsim 0.65$.
This model is not compatible with the scenario, often discussed in the literature,
in which the new heavy gauge bosons have the same coupling to the 
SM currents as the SM gauge bosons. 
At small $\varepsilon$ (small $k$), where the large-$N_T$ approximations
should not be trusted, the coupling grows linearly with $k \propto \varepsilon^2\propto  N_T$.
This is consistent with the expectation from QCD-like models that, 
holding the techni-$\rho$ mass
fixed, the decay constants grow with $N_T$.
Going to large $\varepsilon^2\propto N_T$, this behavior changes smoothly, and
the couplings decrease with $N_T$, consistently with the 
large-$N_T$ picture itself, and the narrow-width approximation. 

The cubic coupling between gauge bosons  can be read off the 
Yang-Mills action
\beqs
{\cal S} &=&\int\di^4x\int_{L_0}^{L_1}\di z\frac{L}{z}\left(1+D\delta(z-L_0)\right)\,g\epsilon^{abc}\,\partial_{\mu}W_{\nu}^aW^{b\,\mu}W^{c\,\nu}\,,
\eeqs
with the replacement $W^3=(g^{\prime}V+g A)/\sqrt{g^2+g^{\prime\,2}}$.
The  relevant cubic couplings $g_{a^{\prime}bc}$ for the computation of the 
techni-$\rho$ decay rates $\Gamma[a^{\prime}\rightarrow bc]$
into longitudinally polarized SM gauge  bosons are:
\beqs
g_{Z^{\prime}WW} &=& \frac{g^2}{\sqrt{g^2+g^{\prime\,2}}}
\int_{L_0}^{L_1}\di z\frac{L}{z}\left(1+D\delta(z-L_0)\right)\,
v_a(z,M_Z^{\prime})\,v_{+}(z,M_W)^2\,,\\
g_{\gamma^{\prime}WW} &=& \frac{gg^{\prime}}{\sqrt{g^2+g^{\prime\,2}}}
\int_{L_0}^{L_1}\di z\frac{L}{z}\left(1+D\delta(z-L_0)\right)\,
v_v(z,M_{\gamma}^{\prime})\,v_{+}(z,M_W)^2\,,\\
g_{W^{\prime}WZ} &=& \frac{g^2}{\sqrt{g^2+g^{\prime\,2}}}
\int_{L_0}^{L_1}\di z\frac{L}{z}\left(1+D\delta(z-L_0)\right)\,
v_{+}(z,M_W^{\prime})\,v_{+}(z,M_W)\,v_a(z,M_Z)\,,\\
\eeqs
with the expressions defined above for the wave functions.

More comments are  in order, about the different behavior of the cubic couplings.
All of them  vanish in the limit in which there is no electro-weak symmetry breaking,
and vanish asymptotically at larg-$N_T$ (large $\varepsilon^2$).
On one hand, this implies that $g_{\gamma^{\prime}WW} < M^2_W/M^2_{W^{\prime}}$.
On the other hand, it implies that it is not possible to approximate the bulk profiles 
with the expression valid for the unbroken case, and it is hence necessary to 
perform the integrations numerically.
A precise computation of these couplings is somewhat problematic also from a more
rigorous point of view: they arise at 
the next-to-leading order in the large-$N_T$ expansion, and are hence sensitive to 
higher-order effects that have not been included in the action under study here.
The (numerical) estimate produced here should not be taken too literally,
because affected by large intrinsic uncertainties. For the purposes of the present analysis 
it is enough to show that, in the
interesting region of the parameter-space, these interactions are subleading and 
can be neglected, as will be shown in the next section.

\section{Techni-$\rho$ decay rates.}
 
The techni-$\rho$ resonances 
 decay both via new strong interactions (to longitudinally polarized SM gauge bosons) and
via weak-interactions (due to the direct coupling to SM fermions).
In all of this analysis, the basic assumption is that there are no other new states lighter than
the techni-$\rho$'s.

Explicitly, the decay rates of the heavy gauge bosons 
into a final state of two light gauge bosons read
\beqs
\Gamma[Z^{\prime}\rightarrow W W]&\simeq&
\frac{g_{Z^{\prime}W W}^2M_{Z^{\prime}}}{48\pi}\,\frac{M_{Z^{\prime}}^4}{M_W^4}\,,\\
\Gamma[\gamma^{\prime}\rightarrow W W]&\simeq&
\frac{g_{\gamma^{\prime}W W}^2M_{\gamma^{\prime}}}{48\pi}\,\frac{M_{\gamma^{\prime}}^4}{M_W^4}\,,\\
\Gamma[W^{\prime}\rightarrow W Z]&\simeq&
\frac{g_{W^{\prime}W Z}^2M_{W^{\prime}}}{48\pi}\,\frac{M_{W^{\prime}}^4}{M_W^2M_Z^2}\,,
\eeqs
up to corrections of $O(M_{W,Z}^2/M_{W^{\prime}}^2)$, where the effective couplings
have been computed in the previous section.
This expression can be directly compared, via the Goldstone boson equivalence
theorem, to the result in a QCD-like theory
$
\Gamma[\rho\rightarrow\pi\pi]\,=\,\frac{M_{\rho}}{48\pi}g_{\rho\pi\pi}^2\,
$.

The decay rates to SM fermions can be read from the leading order 
expressions for the decay rates of the SM gauge bosons, appropriately modified to 
include the decay to top, the heavier mass and the new effective coupling.
For the massive $W^{\prime}$ one can 
compare to the result in the SM in the limit of vanishing top mass:
\beqs
\Gamma_W^t&=&12\,\times\,\frac{g_4^2M_W}{48\pi}\,\simeq\,2.7\,{\rm GeV}\,,
\eeqs
which differs from the experimental value of $2.1$ GeV by the top-bottom final state.
Rescaling:
\beqs
\Gamma[W^{\prime}\rightarrow f\bar{f}]&=&
\left(\frac{g_{c.c.}^{\prime}}{g_{c.c.}}\right)^2\frac{M_W^{\prime}}{M_W}\Gamma_W^t\,.
\eeqs

For the $Z$ boson, 
at the tree level, for vanishing masses of the final-state fermions, 
summing over the three families (top included) would yield:
\beqs
\Gamma_Z^{t}
&\equiv&24\,\left(\frac{g_4^2+g_4^{\prime\,2}}{96\pi}M_Z\right)
 \left(1-2\sin^2\theta_W+\frac{8}{3}\sin^4\theta_W\right)\simeq\,2.8\,{\rm GeV}\,,
\eeqs
which is bigger than the measured width
$
\Gamma_Z\simeq 2.5 \, {\rm GeV}\,,
$
roughly by the top contribution.
Hence, the partial width of the $Z^{\prime}$ to SM fermions:
\beqs
\Gamma[Z^{\prime}\rightarrow f\bar{f}]&=&\left(\frac{g_{n.c.}^{\prime}}{g_{n.c.}}\right)^2\,\frac{M_{Z^{\prime}}}{M_Z}\,\Gamma_Z^t\,.
\eeqs
 Finally, for  the massive photon
\beqs
\Gamma[\gamma^{\prime}\rightarrow f \bar{f}]
&=&8\,\frac{4e_4^{\prime\,2}M_{\gamma^{\prime}}}{48\pi}
\,=\,\frac{8\alpha}{3} R M_{\gamma^{\prime}}
\,.
\eeqs

\begin{figure*}[ht!]
\begin{center}
\includegraphics[width=0.8\linewidth]{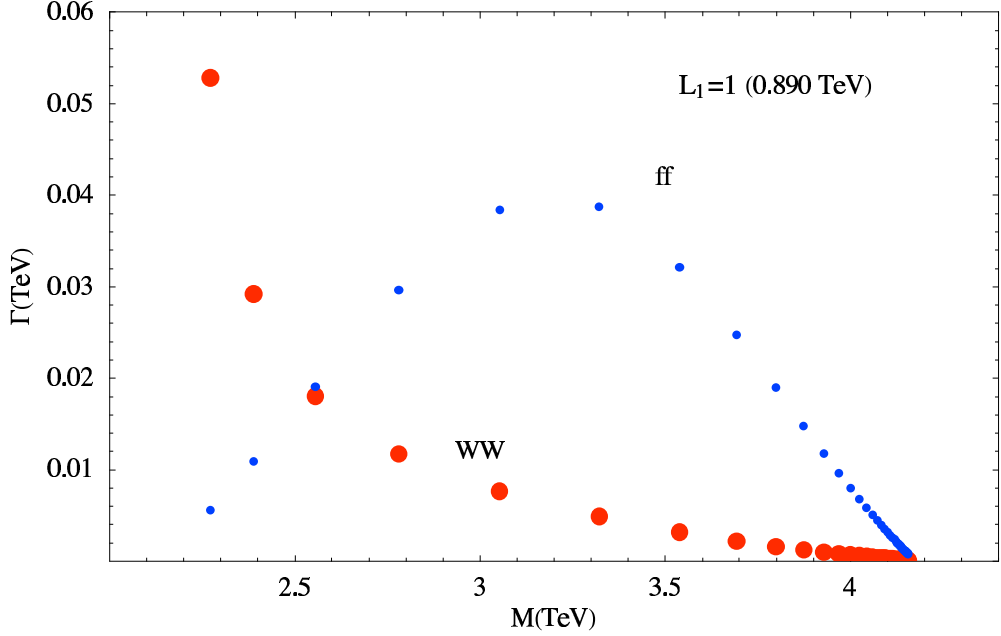}
\caption{Numerical results for 
the partial decay rates $\Gamma[\gamma^{\prime}\rightarrow f\bar{f}]$
and $\Gamma[\gamma^{\prime}\rightarrow W^{+}W^{-}]$ as a function of $M=M_{\gamma^{\prime}}$,
for $L_1=(0.89$ TeV$)^{-1}$.
\label{Fig:decays}}
\end{center}
\end{figure*}

The decay either 
to SM gauge bosons or to SM fermions  dominates the total width of the heavy states,
depending on the value of $\varepsilon$ (or of $k$, see Fig.~\ref{Fig:decays}).
At small $k$ (i.~e. small $\varepsilon^2\propto N_T$), the dominant decay channel is the one
with longitudinal gauge bosons in the final state (as would be the case in a
QCD-like theory). 
This decay rate is a monotonically decreasing function of $\varepsilon^2$,
it scales as $\Gamma\propto 1/N_T$ and hence the partial width decreases
fastly going toward the large-$N_T$ regime, while it is asymptotically large 
at small-$N_T$.
 
 The decay rate to SM fermions is not a monotonic function of $\varepsilon$,
 as clear from the effective coupling $R$ (see figure~\ref{Fig:coupling}).
Starting at intermediate values of $\varepsilon \gsim 1/2$, the decay rate 
of the spin-1 resonances is 
dominated by two-fermion final states. 
At large values of $\varepsilon$, both decay rates fall down as $1/\varepsilon^2$,
but still the two-fermion final state dominates.
Validity of the perturbative approach followed here requires that $\varepsilon>1/2$.
In this range, decay into fermions is always the dominant channel.
The experimental signature of this model reduces to 
that of a model with a complete set of quasi-degenerate heavy copies of the SM
gauge bosons, which decay to the same channels as the SM gauge bosons.

This is the most intriguing and important result of this analysis.
It means that, experimentally, the signature to look at is a relatively clean one,
with a two-body final state that can be explored even in early stages of the LHC
program. In this case, the theoretical control over the estimates is good,
since the one-point functions relevant here are better controlled that the
three-point functions relevant for decays into SM gauge bosons, 
and since at large $\varepsilon$ all the resonances become narrow. 

\section{Two-lepton processes at the LHC.}
For the discovery of $Z^{\prime}$ and $\gamma^{\prime}$ at the LHC I focus
on $pp\rightarrow \mu^{+}\mu^{-}\,+\,X$.  Identical considerations apply also to
 final states with two electrons, which experimentally is a more favorable channel~\cite{AtlasTDR}. 
Decay to $\tau$ and to hadrons, conceptually very similar at
the parton level, would require a more refined analysis of the hadronization process.
At the parton level, this process is described by $q\bar{q}\rightarrow \mu^{+}\mu^{-}$.
It has only  $s$-channel contribution, and can be written by generalizing the
SM tree-level results~\cite{BP}
\beqs
\hat{\sigma}_{q\bar{q}}(\hat{s})\,
\equiv\,\sigma(q\bar{q}\rightarrow\gamma,Z\rightarrow\mu^{+}\mu^{-})&=&
\frac{\hat{s}}{48\pi}\sum_{A,B}\left|G_{AB}(\hat{s})\right|^2\,,
\eeqs
where
\beqs
G_{AB}(\hat{s})&=&Q^{(q)} e_4^2 \left(\frac{}{}P_{\gamma}(\hat{s})+R\,P_{\gamma^{\prime}}(\hat{s})\right)\,+\,
\left(g_4^2+g_4^{\prime\,2}\right)g^{(q)}_Ag^{(\mu)}_B\left(\frac{}{}P_Z(\hat{s})+R\,P_{Z^{\prime}}(\hat{s})\right)
\label{Eq:amplitude}\,.
\eeqs
In this expression $Q^{(f)}$ is the electric charge, and the couplings of the 
SM elementary fermion $f$ are defined by
\beqs
\left\{
\begin{array}{ccc}
g^{(f)}_L&=&T^{3\,(f)}-Q^{(f)}\sin^2\theta_W
\,\\
g_R^{(f)}&=&-Q^{(f)}\sin^2\theta_W\,
\end{array}\right.
\,,
\eeqs
while the propagators for the $V$ gauge boson by
\beqs
P_V(s)&=&\frac{1}{s-M_V^2+iM_V\Gamma_V}\,.
\eeqs

In order to compute the physical cross-sections, define, for $a,b=u,d,s,c,b$,
 \beqs
 \ell_{a,\bar{b}}\left(\hat{s},s\right)&\equiv&\frac{1}{s}
 \int_{\hat{s}/{s}}^1\frac{\di \eta}{\eta}\,\left[\phi_a\left(\eta\right)
 \phi_{\bar{b}}\left(\frac{\hat{s}}{\eta s}\right)\,+\, \phi_{\bar{b}}\left(\eta\right)\phi_a\left(\frac{\hat{s}}{\eta s}\right)
\right]\,,
 \eeqs
 where at the LHC $\sqrt{s}=14$ TeV,  $\hat{s}$ is the c.m. energy of the 
parton-parton  (or equivalently muon-muon) system and
 the $\phi_a$ are parton distribution fuctions as defined by the CTEQ collaboration~\cite{cteq5},
 evaluated at the relevant partonic scale.
The cross-section is
 \beqs
 \frac{\di \sigma(pp\rightarrow \mu^{+}\mu^{-}\,+\,X)}{\di\hat{s}}\left(\hat{s},s\right)
 &=&\frac{1}{3}\sum_a\hat{\sigma}_{a\bar{a}}(\hat{s})\ell_{a\bar{a}}(\hat{s},s)\,,
\eeqs
where the factor of 3 comes from the average 
over initial color states.

The analog process involving a charged vector state is described by:
  \beqs
 \hat{\sigma}_{u\bar{d}}(\hat{s})
 &=&
 \frac{g_4^4|V_{ud}|^2\hat{s}}{192\pi}\left|\frac{1}{\hat{s}-M_W^2+i\Gamma_WM_W}
 \,+\,\left(\frac{g_{c.c.}^{\prime}}{g_{c.c.}}\right)^2\,
 \frac{1}{\hat{s}-M_{W^{\prime}}^2+i\Gamma_{W^{\prime}}M_{W^{\prime}}}\right|^2\,.
 \eeqs

 The $pp$ cross-section is hence:
 \beqs
 \frac{\di
\sigma(pp\rightarrow \mu^{+}\nu_{\mu}\,+\,X)}{\di \hat{s}}\left(\hat{s},s\frac{}{}\right)&=&
\frac{1}{3}\sum_{u,d}\hat{\sigma}_{u\bar{d}}(\hat{s})\ell_{u\bar{d}}(\hat{s},s)\,.
 \eeqs
 
 Unfortunately, the longitudinal energy of the neutrino is not a measurable quantity.
In terms of the transverse mass $m_T$
 of the lepton/neutrino pair:
  \beqs
\frac{\di\hat{\sigma}_{u\bar{d}}(m^2_T,\hat{s})}{\di m^2_T}&=&
\hat{\sigma}_{u\bar{d}}(\hat{s})\,\frac{3}{4\hat{s}}\,\frac{2-m_T^2/\hat{s}}{\sqrt{1-m_T^2/\hat{s}}}
\eeqs
and the observable is the integral over $\hat{s}$ of the differential cross-section
\beqs
 \frac{\di
\sigma(pp\rightarrow \mu^{+}\nu_{\mu}\,+\,X)}{\di m_T^2\,\di \hat{s}}\left(m_T^2,\hat{s},s\frac{}{}\right)&=&
\frac{1}{3}\sum_{u,d}\frac{\di \hat{\sigma}_{u\bar{d}}(m_T^2,\hat{s})}{\di m_T^2}\,\ell_{u\bar{d}}(\hat{s},s)\,.
\eeqs

 \section{Approximate analysis and discussion}
\begin{figure*}[ht]
\begin{center}
\includegraphics[width=0.8\linewidth]{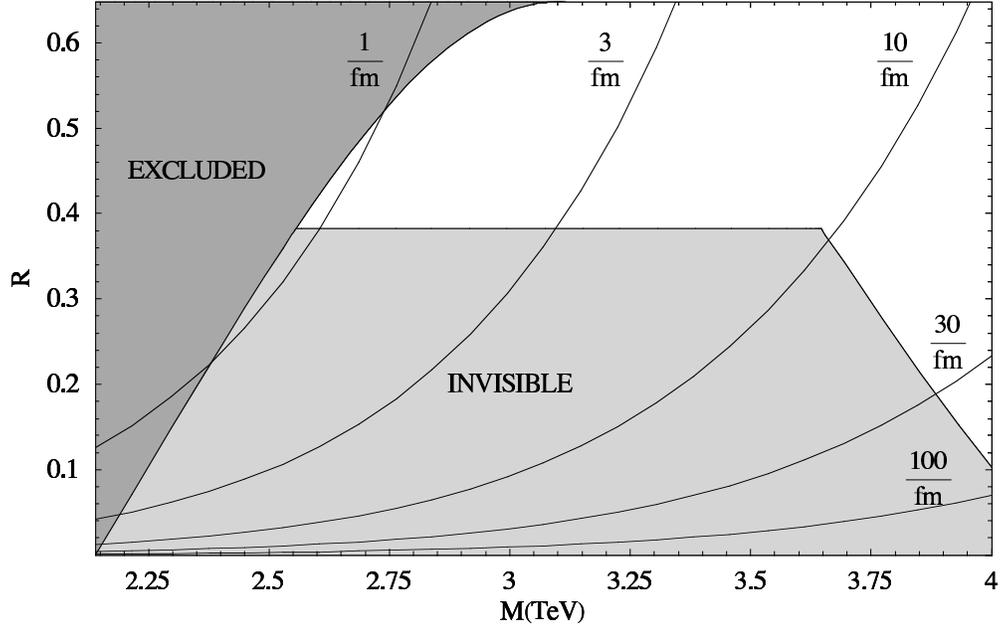}
\caption{LHC exclusion/discovery reach from $pp\rightarrow \mu^{+}\mu^{-}\,+\,X$
as a function of $M\simeq M_{\gamma^{\prime}}\simeq M_{W^{\prime}}
\simeq M_{Z^{\prime}}$ and of $R=|e_4^{\prime}/e_4|^2$. The curves are 
obtained by requiring $10$ events, for integrated luminosity of $1,3,10,30,100$
 fm$^{-1}$. The darkest region is excluded by indirect limits from $\hat{S}<3\times 10^{-3}$.
The light-grey shaded region is allowed by precision data only for
$\varepsilon < 0.5$, so that the dominant decay mode of
the techni-$\rho$'s is into longitudinally polarized SM gauge bosons (and hence
the signal in SM fermions strongly attenuated, or {\it invisible}) and the
large-$N_T$ approximations used in the analysis performed here do not hold.
\label{Fig:exclusions10}
}
\end{center}
\end{figure*}

 Requiring that $N(pp\rightarrow \mu^{+}{\mu}^{-}\,+\,X)>10$,
it is possible to study the range of parameter space explored by the LHC as
a function of its integrated luminosity. Fig.~\ref{Fig:exclusions10} shows a preliminary study,
with no background included and assuming perfect detector efficiency.
No QCD correction ($K$-factor) is added.

\begin{figure*}[ht]
\begin{center}
\includegraphics[width=0.4\linewidth]{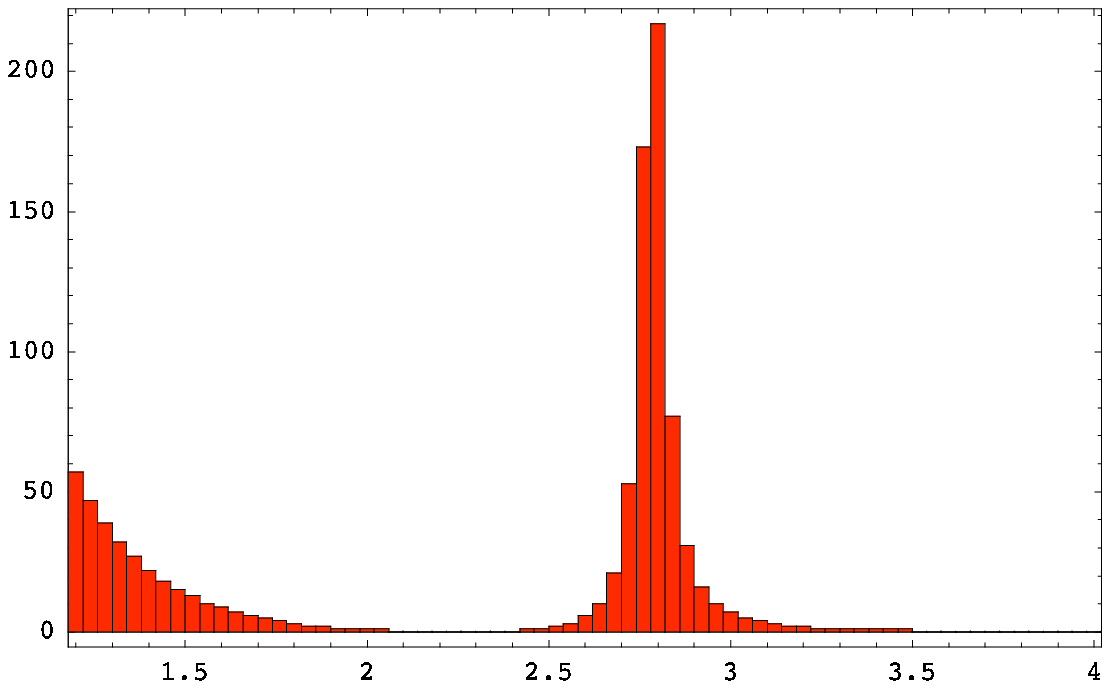}
\includegraphics[width=0.4\linewidth]{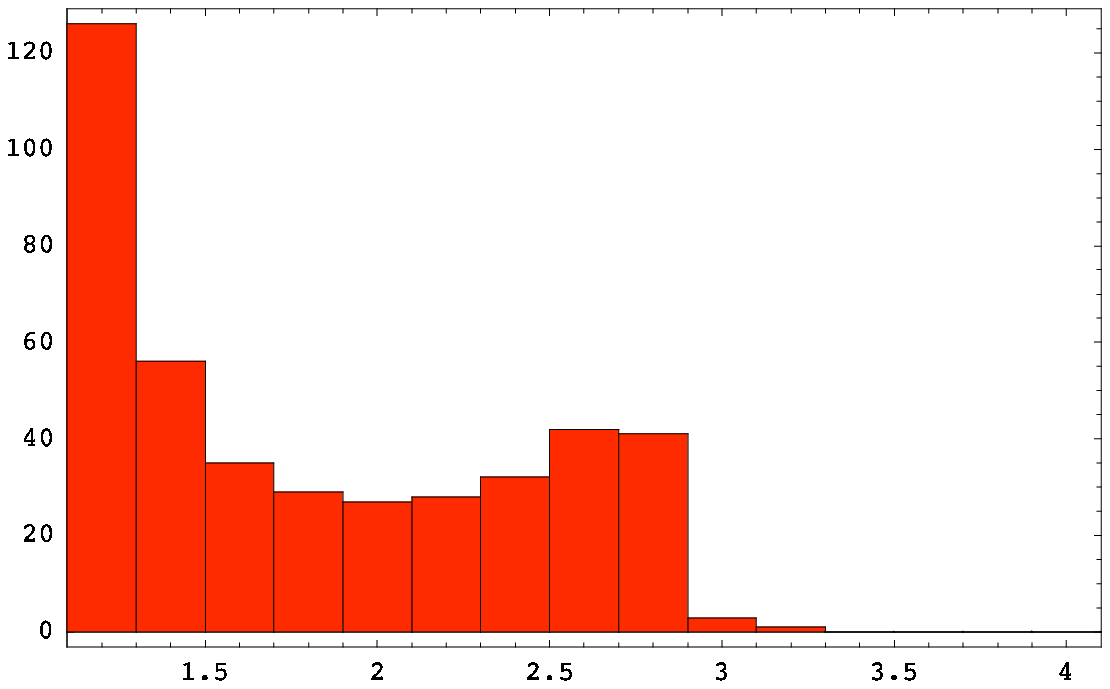}
\includegraphics[width=0.4\linewidth]{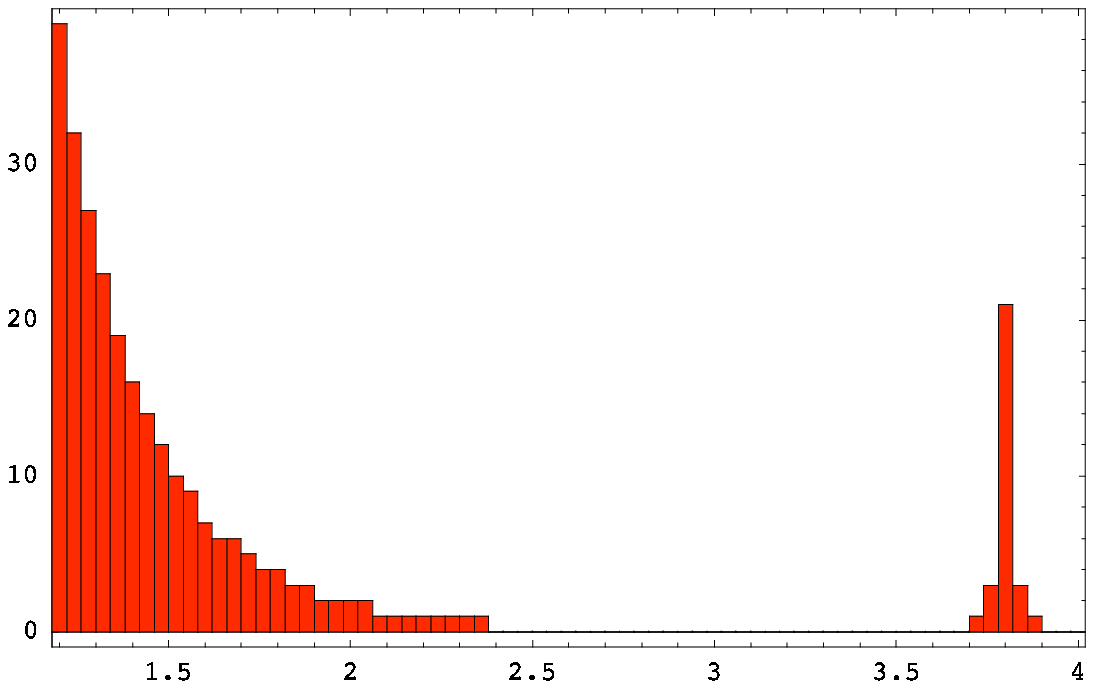}
\includegraphics[width=0.4\linewidth]{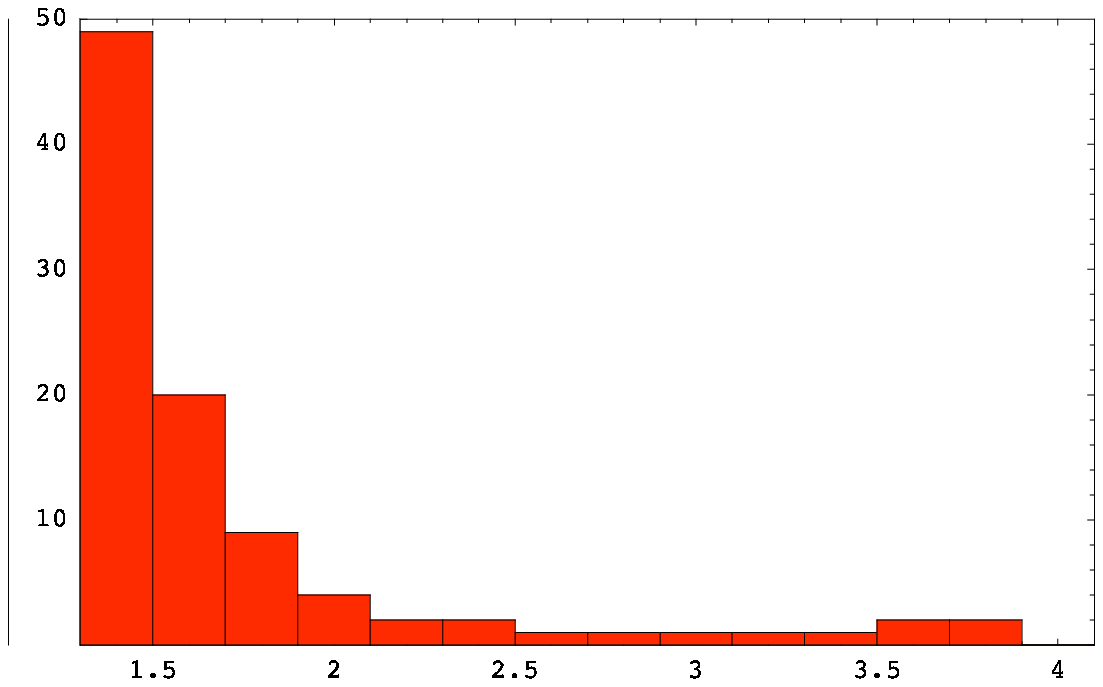}
\caption{Number of events per bin expected at the LHC for integrated luminosity of 
$100$ fm$^{-1}$ and for $L_1=1/0.89$ TeV$^{-1}$, as a function of 
the recostructed $\sqrt{\hat{s}}$ for $\mu^{+}\mu^{-}$ final state
(left diagrams) and of $m_T$ for $\mu^{+}\nu_{\mu}$ (right diagrams).
 Upper diagrams for $\varepsilon=0.6$
(or equivalently for $M_{\gamma^{\prime}}=2.78$ TeV and $R= 0.55$).
Lower diagrams for  for $\varepsilon=1.1$
(or equivalently for $M_{\gamma^{\prime}}=3.80$ TeV and $R= 0.25$).
\label{Fig:fakedata}
}
\end{center}
\end{figure*}

In order to study the properties of these resonances, large statistics is required.
Fig.~\ref{Fig:fakedata} shows the expected distribution of the number of events
for two examples of allowed points in the parameter space, both at the boundary of the exclusion region from precision electro-weak data $L_1=1/0.89$ TeV$^{-1}$ and for integrated luminosity of
$100$ fm$^{-1}$.
The background shown here is the contribution to the same parton-level process
of the SM photon, $W$ and $Z$. Accidental backgrounds from higher order processes 
are not included, perfect efficiency and purity of the signal, and energy resolutions,
have been used (even for the missing energy), and no kinematical cuts  applied. 

The simple numerical study performed here and shown in Fig.~\ref{Fig:exclusions10} and Fig.~\ref{Fig:fakedata}
shows already some interesting result.
Detection of the neutral vectors is much easier, because the complete momentum of the
final state leptons can be reconstructed, and the resonances are rather narrow.
Although none of the potential experimental limitations are included here,
most of the allowed parameter-space can be explored already with $10-30$ fm$^{-1}$.
Kinematical cuts should not reduce the statistics, due to the heavy masses,
and the combination of muon and electron final-states, besides  improving the statistics, 
should  help with energy calibration systematics.

Establishing (or excluding) this model, requires to distinguish in the $\mu^{+}\mu^{-}$ peak
the contribution of $\gamma^{\prime}$ and $Z^{\prime}$, and a precise measurement
of the couplings to up and down quarks,  which are going to be 
quite difficult, because of the large mass of the vector states. 
For the discovery of the $W^{\prime}$,
and the measurement of the helicity of its couplings~\cite{Rizzo},
the smearing of the $m_T$ 
distribution represents a serious limitation.
To some extent, this can be improved with higher luminosity, and
adding together the $\mu$, $e$ and possibly $\tau$ final states.

 \section{Conclusions}
   
I constructed a viable model for dynamical electro-weak symmetry breaking
 by using the five-dimensional language of
the AdS/CFT correspondence. I confronted the model with
the indirect constraints from precision electro-weak measurements,
and identified a large section of the two-dimensional parameter space that
is still allowed and will be directly tested at the LHC even with relatively low luminosity.

 The model is constructed so as to shares some of the basic properties
of more traditional walking technicolor models, but differs from them in two
fundamental ways. First of all, the validity of the perturbative expansion used
in the five-dimensional construction requires to assume that the underlying 
dynamics be well approximated by the large-$N_T$ regime of some fundamental
$SU(N_T)$ theory, and only in that regime the results are reliable.
The parameter controlling this expansion is $\varepsilon$,
and the analysis is performed assuming that $\varepsilon\gsim 1/2$.
Second, the EFT construction used here treats as separate, independent parameters
the scales of confinement of the underlying strong dynamics and the scale of symmetry breaking.
Electro-weak precision constraints are avoided by assuming a (very moderate) 
hierarchy between the two, the confinement scale being related to the
IR cut-off of the conformal sector $L_1\lsim 1/0.89$ TeV$^{-1}$.
This is enough to suppress $\hat{S}$, while $\hat{T}$ does not play an important role.

The spectrum contains no light Higgs, but  a tower of 
spin-1 states with quantum numbers identical to the SM gauge bosons
are the signature to be looked for at the LHC.
The states in the first excited level, denoted by $\gamma^{\prime}$, $W^{\prime}$ and
$Z^{\prime}$, are going to have degenerate masses in the few TeV range.
The decay to longitudinally polarized SM gauge bosons,
which is experimentally challenging, is important (or dominant)
only at small values of $\varepsilon$, i.~e. in the small-$N_T$ regime,
which cannot be described reliably within the approach followed here.
Most important, in the regime in which the control over the EFT is good,
at large values of $\varepsilon$ (large-$N_T$ limit),
the spin-1 states decay mostly into SM fermions via direct coupling to the SM currents.
An absolute upper bound for this coupling is found, insuring that the 
the width is always small, and that perturbative treatment of the relevant 
production and decay rates is accurate.

The search for these new states can be done with traditional analysis methods
used for the search of new heavy gauge bosons with SM-like couplings.
The neutral states can be discovered already with 10 fm$^{-1}$ integrated luminosity
in the $2.5-4$ TeV mass range, provided the parameter $R$, controlling the 
 coupling to the SM currents compared to that of the SM gauge bosons, is not too small.
 Measurement of $R$ and $M_{\gamma^{\prime}}$ are enough to determine the two 
 free parameters $\varepsilon$ and $L_1$ of the EFT.
In order to distinguish this model from a generic $Z^{\prime}$ scenario, larger luminosity
must be cumulated, so as to discover also the charged $W^{\prime}$, distinguish $Z^{\prime}$ and $\gamma^{\prime}$, and show that the masses
and the ratios of couplings $R$ are approximately the same for all the new states. 
Distinguishing the $\gamma^{\prime}$ from the $Z^{\prime}$ 
and establishing the nature of the $W^{\prime}$ can be challenging,
in view both of the largish masses and of the upper limit $R\lsim 0.65$,
but should be possible over a significant part of the parameter space,
once the design luminosity for the LHC is reached.

\vspace{1.0cm}
\begin{acknowledgments}
I would like to thank Stephen Ellis, Matthew Strassler and Wu-Ki Tung
for useful discussions. 
This work is partially supported by the Department of Energy grant
DE-FG02-96ER40956.
\end{acknowledgments}


\end{document}